\documentclass[manuscript]{aastex}
\usepackage{graphicx}
\usepackage{color}

\shorttitle{Flare risk signatures}
\shortauthors{Kors\'os et al.}

\begin{document}

\title{On flare predictability based on sunspot group evolution}

\author{M. B. Kors\'os\altaffilmark{1,2}, A. Ludm\'any\altaffilmark{1}, R. Erd\'elyi\altaffilmark{1,2} and T. Baranyi\altaffilmark{1}}
\altaffiltext{1}{Debrecen Heliophysical Observatory (DHO), Research Centre for Astronomy and Earth Sciences, Hungarian Academy of Science, 4010 Debrecen, P.O. Box 30, Hungary}
\altaffiltext{2}{Solar Physics \& Space Plasma Research Center (SP2RC), University of Sheffield, Hounsfield Road, S3 7RH, UK}
\email{[korsos.marianna; baranyi.tunde; ludmany.andras]@csfk.mta.hu}
\email{robertus@sheffield.ac.uk}

\begin{abstract}

The forecast method introduced by \cite{korsos2014} is generalised from the horizontal magnetic gradient ($G_{M}$), defined between two opposite polarity spots, to all spots within an appropriately defined region close to the magnetic neutral line of an active region. This novel approach is not limited to searching for the largest $G_{M}$ of two single spots as in previous methods. Instead, the pre-flare conditions of the evolution of  spot groups is captured by the introduction of the {\it weighted} horizontal magnetic gradient, or $WG_{M}$. This new proxy enables the potential of forecasting flares stronger than M5. The improved capability includes (i) the prediction of flare onset time and (ii) an assessment whether a flare is followed by another event within about 18 hours. The prediction of onset time is found to be more accurate here. A linear relationship is established between the duration of converging motion and the time elapsed from the moment of closest position to that of the flare onset of opposite polarity spot groups. The other promising relationship is between the maximum of the $WG_{M}$ prior to flaring and the value of $WG_{M}$ at the moment of the initial flare onset in the case of multiple flaring. We found that  when the $WG_{M}$ decreases by about 54\%, then there is no second flare. If, however, when the $WG_{M}$ decreases less than 42\%, then there will be likely a follow-up flare stronger than M5. This new capability may be useful for an automated flare prediction tool. 
\end{abstract}


\section{Introduction}

The endeavour to establish reliable flare forecast methods has resulted in numerous attempts to identify promising physical quantities, proxies, features and behavioural patterns to predict an imminent flare \citep[see e.g.][]{sawyer1986,hochedez2005,benz2008}. The ultimate task is to construct a diagnostic tool that determines the unique conditions leading to flaring events, triggered by magnetic reconnection \citep[see e.g.][]{yamada2010}. The empirical study presented here focuses on features at the solar surface, namely investigating the pre-flare dynamics of sunspot groups. 

Most of the attempts that developed flare forecast tools employ suitably defined (and derived) quantities from magnetograms. The majority of previous efforts studied the behaviour of the horizontal gradient of the line-of-sight component of the magnetic field, usually a well-observed property of an active region (AR). By investigating the Solar and Heliospheric Observatory's Michelson Doppler Imager (SOHO/MDI) magnetograms, \citet{schrijver2007} found that energetic flares are connected to the separation line of opposite polarity regions with high magnetic gradient. Further works followed suit, including e.g. \citet{mason2010} who studied the gradient-weighted inversion line length which exhibits a significant increase prior to flares. The maximum horizontal gradient and the length of the neutral line were considered by \citet{cui2006}, \citet{jing2006}, \citet{huang2010} and \citet{yu2010a}. The fractal structure was addressed by \citet{crisc2009}, however, \citet{georg2012} was sceptical about this approach. In the literature, there are a number of other indicators proposed to measure non-potentiality. \citet{leka07} suggested 8 categories of different distributions with an impressive list of 29 types of variables defined on them. All these methods above (and others) have advantages and caveats, however, none are yet accepted as a universal and reliable prediction tool. 

Surprisingly, sunspots themselves are rarely considered as possible holders of information for flare forecast. Arguably, the simplest way to apply sunspot data is to approach non-potentiality by using their McIntosh classification \citep{bloom12}. However, the McIntosh classes are only based on morphology, and they do not contain perhaps the most important information  relevant to the present context, i.e. the distribution of opposite magnetic polarities. Our preceding study (\citealt{korsos2014}, hereafter Paper I) is the first attempt to track the details of the evolution of sunspots (umbrae) in order to identify signatures of flare imminence. Paper I implemented a proxy horizontal magnetic gradient, $G_M$, defined between two {\it single} spots of opposite polarities close to the magnetic neutral line. The pre-flare behaviour of this proxy quantity exhibited characteristic and unique patterns: steep rise, high maximum and a gradual decrease prior to flaring. These properties may yield a tool for the assessment of flare probability and intensity within a 2-10 hour window.

The method developed here allows us to elaborate on some of the most important properties of an imminent flare; its intensity and the onset time. The prediction of intensity is found to be more reliable, as a linear relationship has been found between the pre-flare maximum of $G_M$ and the peak intensity emitted in the 1-8~\AA\ range, according to Geostationary Operational Environmental Satellite (GOES) x-ray measurements\footnote{http://www.ngdc.noaa.gov/stp/satellite/goes/dataaccess.html}. This result may be considered to be an indicator of the existing relationship between the proxies of free and released energies. Next, the onset time prediction is found to be somewhat less precise, as the most probable time of flare onset is  between 2 and 10 hours after the $G_{M}$ maximum. The aim of the present work is to find more reliable forecasting methods for the onset time as well as the likelihood of consecutive flaring.

\section{Method of Examination}

The empirical basis of this study, similar to Paper I, is the SDD\footnote{http://fenyi.solarobs.unideb.hu/SDD/SDD.html} (SOHO/MDI-Debrecen Data) sunspot catalogue, the most detailed of its kind covering the years of MDI operations (1996-2010). In addition, the intensity peaks of the examined flares the GOES solar flare database\footnote{
http://www.ngdc.noaa.gov/stp/satellite/goes/dataaccess.htm} is employed. The SDD is efficient for tracking the internal dynamics of spot groups. It contains data on position, area and magnetic field for all spots with a cadence of 1.5 hrs \citep{gyori2011}. 

Let us now consider the involvement of {\it all} magnetic spots in an appropriately selected area at the region of a Polarity Inversion Line (PIL). When a new spot (min. 3 MSH) emerges close (within 40 $\pm 5$ Mm, which is always enough to capture the emergence) to existing spots of opposite polarity, we determine the maximum $G_M$ between the emerging and existing spots. Once a pair of spots of opposite polarities with a maximum $G_M$ is found, we compute the area-weighted centre between them. Next, there is a defined circular area, around this weighted location where $G_M$ is highest, whose diameter is 3$^{\circ}$$\pm 0.5$$^{\circ}$ in Carrington heliographic coordinates. The center of the circle is fixed and spot groups within this area are now monitored. 

We assume that the underlying process driving a flare is a collective one between nearby spots. Therefore, a new proxy parameter, a generalisation of the $G_M$, called the weighted horizontal magnetic gradient ($WG_{M}$) is defined to account for this collective behaviour:

\begin{equation}
WG_{M} = \left | \frac {\sum_{i} B_{p,i}\cdot A_{p,i} - \sum_{j} B_{n,j}\cdot A_{n,j}}{d_{pn}} \right |.   
\label{proxy}
\end{equation}

\noindent
Here, $B$ (determined by $f(A)$  in Paper I) and $A$ denote the mean magnetic field and area of umbra. The indices $p$ and $n$ denote positive and negative polarities, $i$ and $j$ are their running indices in the selected spot cluster and $d_{pn}$ is the distance between the area-weighted centers of two subgroups of opposite polarities in this cluster.

We have analysed 45 single and 16 multiple flare cases which are stronger than M5 between 1996 and 2010. This limitation is based on the findings  of Paper I, i.e. the present method seems to be suitable for energetic flares above M5. Besides some similarities between the behaviour of $G_{M}$ and $WG_{M}$, the current approach results in significant, previously unseen pre-flare behavioural patterns. Let us demonstrate the key features with two randomly selected, but typical, examples.

\begin{figure}
\epsscale{0.9}
\plotone{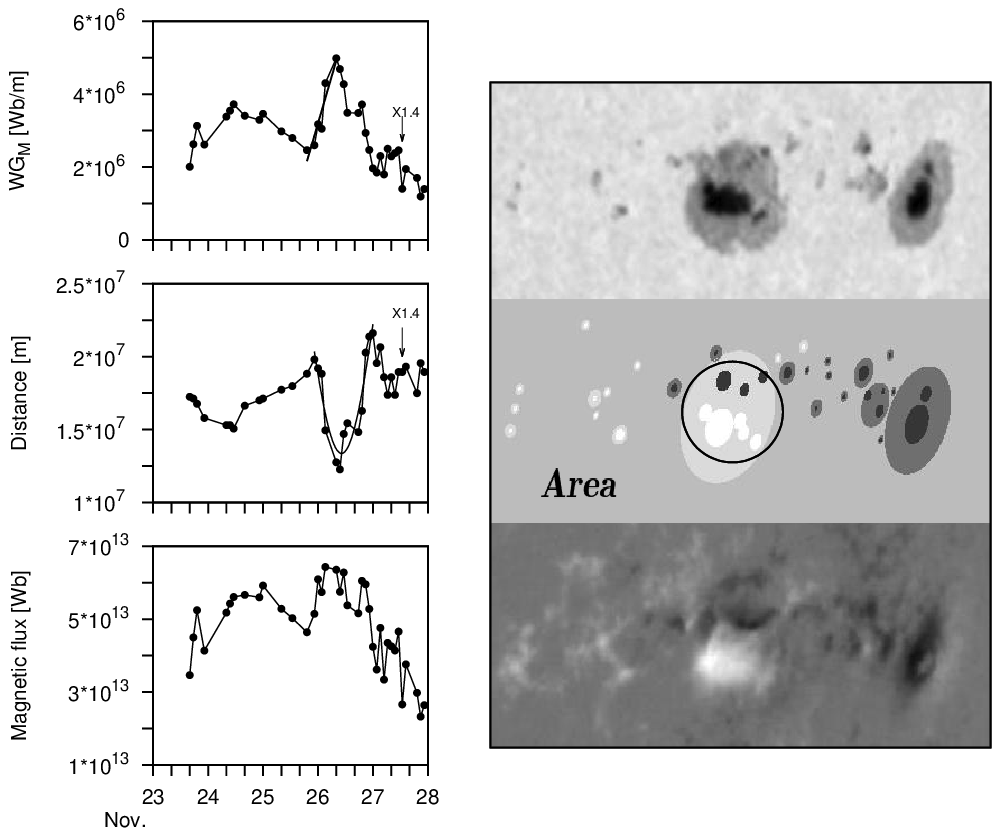} 
\caption{NOAA AR 8771, for Nov. 23-26, 1999. Right column: continuum white-light image (top), reconstruction from SDD (middle), magnetogram (bottom). Left column: variation of $WG_{M}$ (top), distance between the area-weighted centers of the spots of opposite polarities (middle) and unsigned flux of all spots in the encircled area (bottom).  \label{8771}}

\end{figure}

Figure \ref{8771} shows a typical active region, AR 8771, with a single flare. The right-hand panels of Fig. \ref{8771} are: the white light image (top), magnetogram (bottom) and a cartoon reconstructing the AR from the SDD catalogue (middle). The $WG_{M}$ (top left) shows a steep rise and high maximum (called $WG_M^{max}$) followed by its decrease until the flare. However, and most importantly, we found  characteristic and appealing differences from the result of the single spot-pair method, as demonstrated by e.g. the distance diagnostics panel (middle left) of the spot groups of opposite fluxes. This plot contains a conspicuous dip, indicating a duration of converging-diverging motion of the area-weighted centres that seem to be indicative of the next flare for {\it all} cases we investigated. The total unsigned magnetic flux (bottom left) in this part of the AR shows some increase before the flare but it has no special, identifiable characteristic feature. Additionally, we tested a large sample of non-flaring spot groups with opposite polarities to determine whether this behaviour is found in these cases as well. We found no such behaviour. Therefore, the evolution of the distance between the area-weighted centers of spot groups of opposite polarities shows potentials for flare forecast.

Figure \ref{9393} is another typical example, i.e. AR 9393 also examined in Paper I, but now with multiple flare activities linked only to Area 2 because the highest variability of $WG_M$ is in Area 2, while Area 1 and 3 do not show such a property (see in Paper I). A series of flares is considered multiple, if after the occurring flare there is another flare, within an 18-hr window, belonging to the same preceding rising phase of $WG_M$ and at the same time belonging to the {\it same} decreasing phase after $WG_M^{max}$ (see e.g. the X1.4 flare followed by the X20 flare in the upper left panel of Fig. \ref{9393}). The behaviour of $WG_{M}$ is analogous to that of the single spot-pair (compare to $G_M$ of Fig. 2 in Paper I), showing a steep rise, high maximum and a slower decrease prior to the X1.7 single flare (29 March), and, in spite of the limited temporal resolution, arguably a similar pattern is found leading to the follow-up ndependent multiple flares (i.e., X1.4 and X20, 2 April). Next, the distance diagnostics (middle left) panel of the spot groups of opposite fluxes do show similar conspicuous dips as in Fig. \ref{8771}.  Note, these multiple flares are likely unrelated to the single flare as they do not fall within the 18-hour window we propose as a requirement for flares to be connected. Again, this pre-flare behaviour pattern was not identifiable in the distance diagram of the single spot-pair method in Paper I (see its Fig. 2, left column, middle panel). Finally, we could not conclude any special and immediate pre-flare behaviour on a daily time-scale by investigating the variation of unsigned flux within Area 2 (bottom left) because it was decreasing before the X1.7 flare while it was increasing before the X1.4 flare. A statistical study also shows that the temporal variation of the unsigned flux does not exhibit characteristic pre-flare signatures. This, however, does not contradict the findings of \citet{schrijver2007} about a statistical relationship between the likelihood of X- or M-flares and the unsigned flux at SPIL (Strong-gradient Polarity Inversion Line) because selected actual states of ARs were investigated and not the time profile of the examined quantities. 

By the above examples of case studies, we are now encouraged to conduct a statistical study to confirm (or refute) the above findings for all cases of flaring spots with a strength of over M5 available in the SDD catalogue.

\section{Diagnostic potentials with spot dynamics}

We test the proposed diagnostics on a statistical sample by applying the following requirements: Firstly, the examined pre-flare variation is within $\pm 70^{\circ}$ from the central meridian to avoid geometrical foreshortening close to the limb. Next, the flare onset is no further eastward from the central meridian than $-40^{\circ}$ to have sufficient time to follow the development of $WG_{M}$.

In Paper I, a linear relationship was found between the maximum value of $G_{M}$ preceding a flare and the peak intensity of flares. This behaviour is confirmed for $WG_{M}$ by Fig. \ref{invmax}. 45 single flares (crosses, left panel) and 45 single with additional 16 largest of multiple flares (circles, right panel) show a linear relationship between $WG_{M}^{max}$ and the corresponding GOES flare intensity. Here, we restrict the empirical analysis for flares between M5 and X4 classes only.

\begin{figure} 
\epsscale{0.9}
\plotone{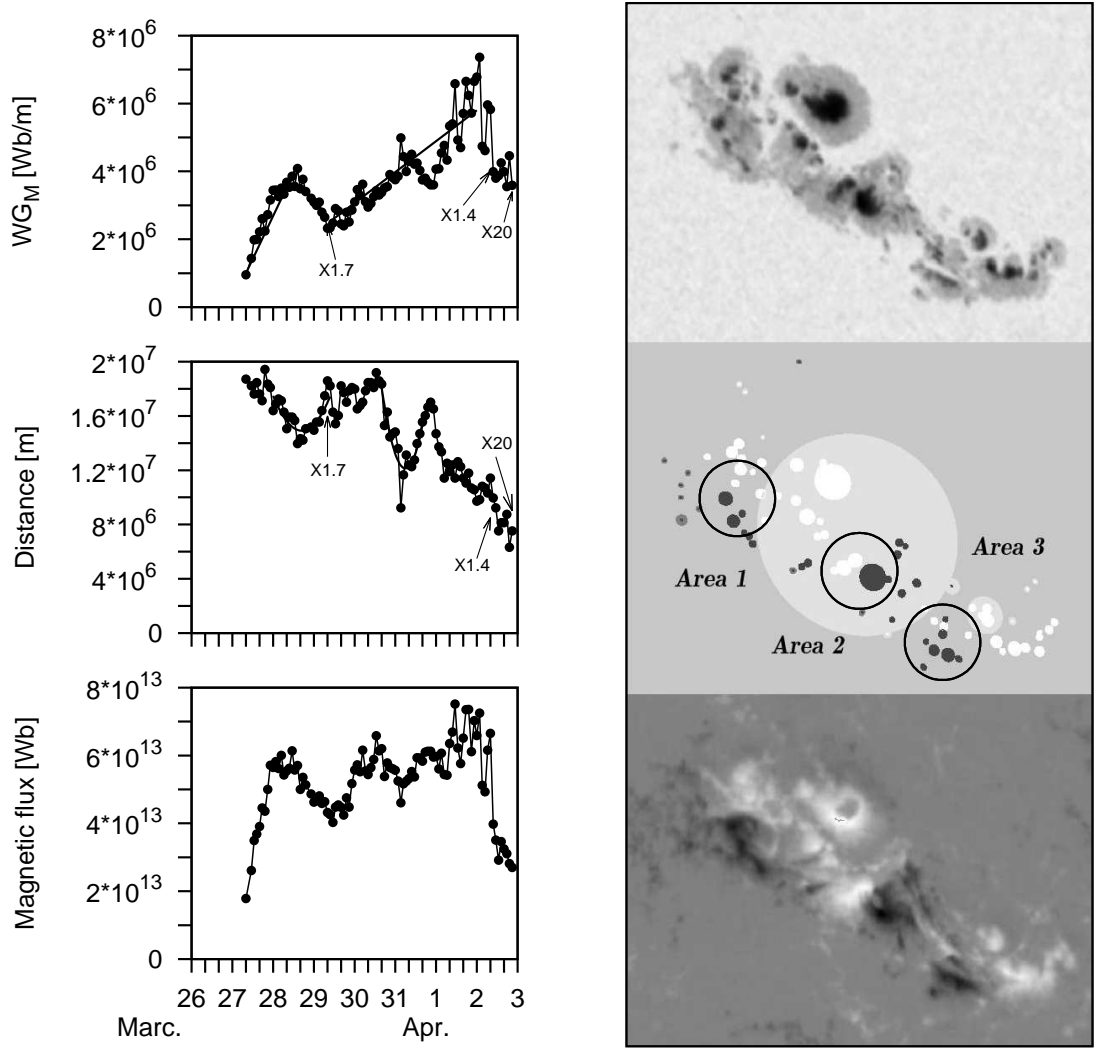} 
\caption{Same as Fig. \ref{8771} but of NOAA AR 9393 with a single (i.e., X1.7) and multiple (i.e., X1.4 and X20) flares, for Mar 26 - 03 Apr 2001.  \label{9393}}
\end{figure}

Next, the new method revealed further important connections, which is conspicuous, in Figs. \ref{8771} and \ref{9393}. This connection is between the durations of converging-diverging motion of the centers of opposite polarities. This intriguing pattern was found in all 61 cases investigated here. The question rises, whether there is a relationship between the duration of the converging motion (the duration from the moment of the first point when the distance began decreasing to the moment of the minimum point of the parabolic curve) and the time elapsed from the moment of minimum distance until the  flare onset (duration of the diverging motion and the follow-up time until the flare onset). To determine these two time intervals for each flare, parabolic curves were fitted to their distance data. For a sample see the top left panel of Fig. \ref{pushpull} which is a parabolic fit to the distance data from the left middle panel of Fig. \ref{8771}, showing the converging-diverging behaviour of this relative motion.

Figure \ref{pushpull} gives further insight into the relation between these intervals by plotting the time from the moment of minimum distance to the flare onset
as a function of the duration of converging motion. First, the upper right diagram depicts the duration of diverging  motion as function of the duration of converging motion for the 45 single (crosses) and 16 multiple (circles) flare cases. Note that the duration of diverging motion  is shorter than the time period from the moment of minimum distance to that of the flare onset (see bottom panels of Fig.~\ref{pushpull}). However, the converging-motion phase and the diverging-motion phase have the same duration. \cite{yamada2010} found similar properties in laboratory reconnection experiments, and called it the "push and pull-mode". The present observations are a confirmation of the laboratory experiment.

The lower diagrams plot the time from the moment of closest position to the flare onset as function of the duration of the converging  motion phase. In the cases of multiple flares, we investigated the time from the moment of closest position to the {\it first} flare onset as function of the duration of the converging motion phase. The left/right panel contains those cases when the spot groups are younger/older than three days at the time of flare onset. The regression lines of the two cases are, surprisingly, different. By estimating the time the magnetic fields younger than three days should be distinguished from the older ones, the relevant formulae are given in the lower panels of Figure \ref{pushpull}. One may be able to estimate a rough onset time of the flare. Note the considerable dispersion. If the study area is younger than three days then about a mere hour is needed to be added to the duration of the corresponding scaled duration of converging motion, where the scale-factor is 1.3 for younger ones, to obtain the flare onset time. However, if the area is older than three days then the scale-factor is 0.85, and one needs to add 12 $\pm 3$ hours to the scaled duration of converging motion.

Next, a relationship is found between the values of $WG_{M}$ at its maximum prior to flaring ($WG_{M}^{max}$) and at the time of flare onset ($WG_{M}^{Flare}$), as visualised in Figure \ref{maxflaregm}. We investigate separately the 45 cases when only a single flare took place after the maximum of $WG_{M}$, and the 16 when more flares erupted after the maximum within an 18-hour window after the occurring flare on the decreasing phase of the $WG_{M}$. The left panel depicts the cases of single energetic flares (crosses). The right panel depicts the first flares (circles) of those ARs where multiple flares are produced. The plots are interpreted as follows: a single flare erupts when the $WG_{M}^{max}$ was less than $5\cdot10^{6}$ Wb/m and the $WG_{M}$ decreases by more than half of the $WG_{M}^{max}$, in the pre-flare phase (for example the X1.4 flare in the NOAA 8771 on Fig.~\ref{8771} and the X1.7 flare in the AR NOAA 9393 on Fig.~ \ref{9393}). This is likely due to the fact that the magnetic energy in the region decreases so significantly during this first flare that there is simply not enough energy left to release another flare. In that case, if the decrease is smaller than half (about $42$\%) by the onset of the first flare, some further flaring may be expected, meaning that the disturbance of this first flare could be forcing opposite polarity fields together in the solar atmosphere leading, for example to the `homologous' flaring that is often observed. On the other hand, the $WG_{M}^{max}$ larger than $5\cdot10^{6}$ Wb/m  seems to be enough in itself to predict a multiple flaring event. We cannot comment yet on further flares (i.e. third, etc.) in the case of multiple flares as the temporal resolution of the SDD catalogue is too coarse.  For an example, see the case of AR NOAA 9393 where an X 1.4 flare was followed by an X 20 one within 12 hours (Fig. \ref{9393}).

\section{Discussion}

In this paper, we present advancements in the classification of pre-flare conditions with an application to flare prediction. 61 cases were investigated in the vicinity of PILs of ARs. We assumed the flare onset to be a collective response to their dynamics.  This assumption needs further investigations both observationally (with higher resolution) and theoretically (e.g. numerical simulations).

\begin{figure} 
\epsscale{1}
\plotone{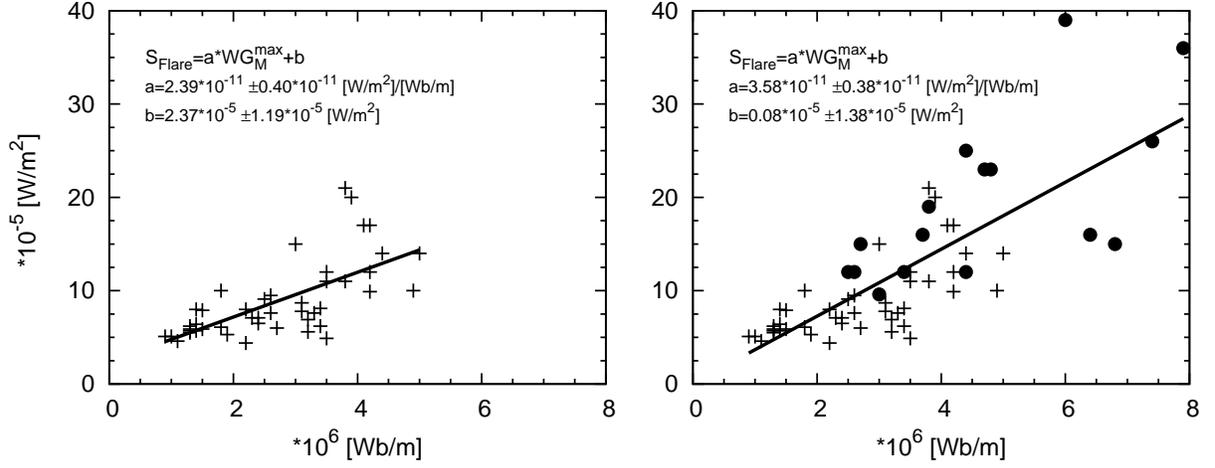}
\caption{GOES intensity of flares as function of the maximum $WG_{M}$. The left diagram shows the intensity of single solar flares (crosses) that occurred after the maximum $WG_{M}$. The right diagram depicts, in addition, the flares (circles) with largest intensity within an 18-hr interval after reaching $WG_{M}^{max}$.}
\label{invmax}
 
\end{figure}

\begin{figure}
\epsscale{1}
\plotone{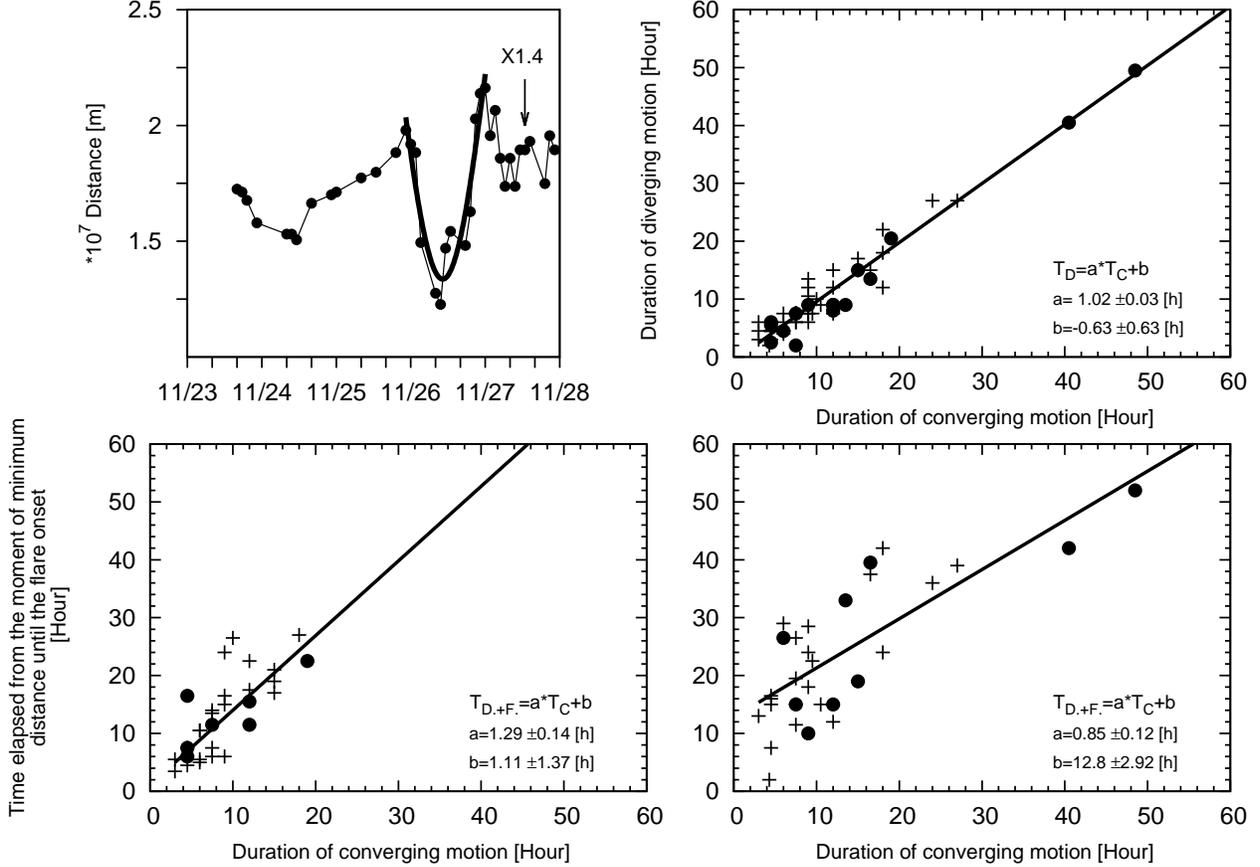}
 \caption{Upper left: The converging-diverging pattern with a parabolic fit to the data for AR 8771.  Upper right: Relationship between the  durations of converging and diverging motion for all 61 flares. Crosses (45)/circles (16) indicate single/first of multiple flares. Please note that apparent number of point may not be 45/16 in the plots because there are overlapping data points. Lower panels: Duration of diverging motion until flare onset as function of duration of the compressing phase of motion of opposite polarities. The study area is younger (left)/older (right) than three days.}

	\label{pushpull}
\end{figure}

\begin{figure}
\epsscale{1}
\plotone{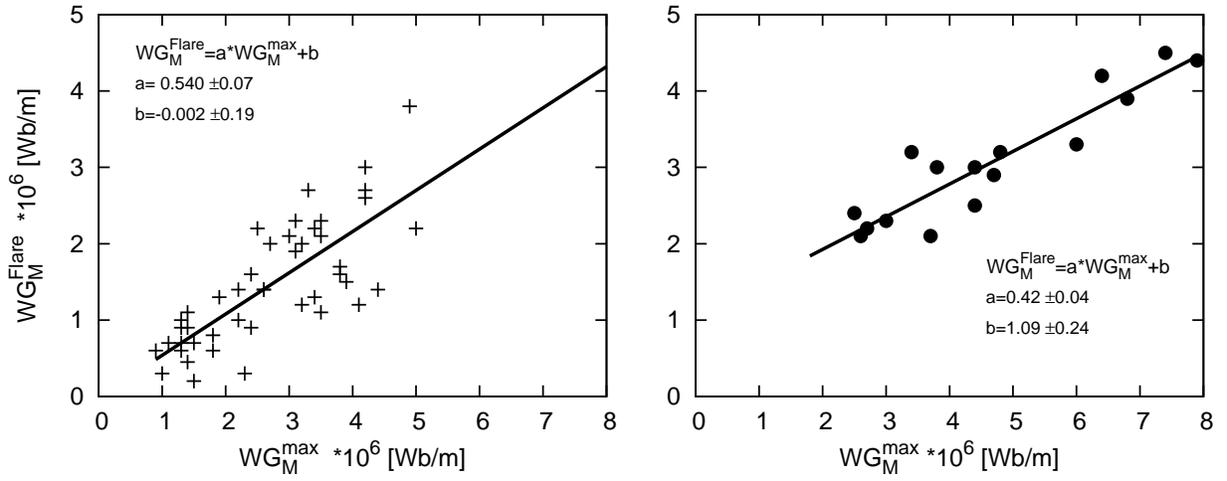} 
 \caption{The weighted horizontal magnetic gradient at flare onset $WG_{M}^{Flare}$ as function of the maximum of the weighted horizontal magnetic gradient prior to flare $WG_{M}^{max}$. Left panel: cases of a single flare. Right panel: first events of multiple flares after $WG_{M}^{max}$.}
\label{maxflaregm}
\end{figure}

First, we found that the pre-flare behaviour of the weighted horizontal magnetic gradient ($WG_{M}$) exhibits similar patterns to those found with the single spot-pair method: steep rise, high maximum and gradual decrease until the flare onset (Figs. \ref{8771}-\ref{9393}). Next, Figure \ref{invmax} corroborates the relationship, reported in Paper I, between the maximum of $G_{M}$ and the intensity of single flares. Here, this relationship is modelled as linear one, however, the dispersion is considerable and theoretical (e.g. numerical) modelling may be necessary to confirm or refute this relation. There may be a yet unknown physical parameter, therefore not accounted for, that would reduce the dispersion. This is the reason of restriction on the currently considered GOES classes. However, the relationship found can be still regarded to be a link between the proxy measures of the free energy and the released energy. A shortfall of the single spot-pair method was that one could not deduce the flare intensity in the case of multiple flares. This is now rectified by the introduction of the weighted horizontal magnetic gradient. The spot-group method is now capable of providing a rough estimate of the expected largest flare intensity from $WG_{M}^{max}$ (Fig.~\ref{invmax}). In addition, this method also gives a better estimate of the expected time of flare onset (Fig.~\ref{pushpull}), and, may be able to predict whether an energetic flare after $WG_{M}^{max}$  is the only flare or further flare events can be expected (Fig.~\ref{maxflaregm}). 

Let us assume that $WG_{M}$ is a proxy of the available non-potential (i.e. free) energy to be released in a spot group. In this case, we may conclude from Fig. \ref{maxflaregm}: (i) if the maximum of the released energy may be over half of the maximum of the accumulated (free) energy, no further energetic flare(s) can be expected; (ii) If the maximum of the released flare energy is less than about $\sim$42\%, further flares are more probable. In short, Fig. \ref{maxflaregm} allows us to track the variation of the energy balance of ARs and to assess the probabilities of consecutive flares and their intensities.

Last but not least we provide some notes on the estimate of the onset time of an imminent flare. Here, its determination is refined. Paper I only presented a statistics that 60\% of observed energetic flares are between 2-10 hrs after the maximum of $G_{M}$. Figure \ref{pushpull}, however, allows a much stronger statement on the expected time of onset due to $WG_{M}$. The figure uncovers the relationship between the duration of the converging motion of opposite polarities (their compression) and the time elapsed between the closest position and flare onset following the diverging motion \citep[see earlier motions][]{yamada2010}. By determining the duration of the converging motion the flare onset can now be assessed for {\it all} cases. We also found that the data points of the motions of younger spot groups have smaller dispersion (left of Fig. \ref{pushpull}).
\section{Acknowledgment}

The research leading to these results has received funding from the European Community's Seventh Framework Programme (FP7/2012-2015) under grant agreement No. 284461 (eHEROES project). MBK and RE are grateful to Science and Technology Facilities Council (STFC) UK for the financial support received. RE would like to thank for the invitation, support and hospitality received from the Hungarian Academy of Sciences under their Distinguished Guest Scientists Fellowship Programme (ref. nr. 1751/44/2014/KIF) that has allowed him to stay three months at the Debrecen Heliophysical Observatory (DHO) of the Research Centre for Astronomy and Earth Sciences, Hungarian Academy of Sciences. RE is also grateful to NSF, Hungary (OTKA, Ref. No. K83133).

\clearpage

\clearpage

\clearpage

\clearpage

\clearpage

\end{document}